\documentclass[prl,twocolumn,longbibliography]{revtex4-1}
\usepackage{epsfig}
\usepackage{graphicx}
\usepackage{amsmath}
\usepackage{natbib}
\usepackage{color}

\newcommand{\n}{\nonumber}
\newcommand{\bn}{\begin{eqnarray}}
\newcommand{\en}{\end{eqnarray}}
\newcommand{\eml}{\end{multline}}
\newcommand{\bml}{\begin{multline}}
\newcommand{\h}{\hspace}

\begin{document}
 \author{Kunal K. Das$^{1,2}$ and Miroslav Gajdacz$^{3,4}$}
 \affiliation{$^{1}$
Department of Physical Sciences, Kutztown University of Pennsylvania, Kutztown, Pennsylvania 19530, USA}
\affiliation{$^{2}$Department of Physics and Astronomy, Stony Brook University, Stony Brook, New York 11794-3800, USA}
 \affiliation{$^{3}$ OFS Fitel, Priorparken 680 Broendby, Copenhagen 2605, Denmark}
  \affiliation{$^{4}$ Department of Physics and Astronomy, Aarhus University, Ny Munkegade 120, 8000 Aarhus C, Denmark}

\date{\today }

\title {Synthetic Gauge Structures in Real Space in a Ring lattice}

\begin{abstract}Emergence of fundamental forces from gauge symmetry is among our most profound insights about the physical universe. In nature, such symmetries remain hidden in the space of internal degrees of freedom of subatomic particles. Here we propose a way to realize and study gauge structures in real space, manifest in external degrees of freedom of quantum states. We present a  model based on a ring-shaped lattice potential, which allows for both Abelian and non-Abelian constructs.  Non trivial Wilson loops are shown possible via physical motion of the system. The underlying physics is based on the close analogy of geometric phase with gauge potentials that has been utilized to create synthetic gauge fields with internal states of ultracold atoms. By scaling up to an array with spatially varying parameters, a discrete gauge field can be realized in position space, and its dynamics mapped over macroscopic size and time scales.\end{abstract}

\maketitle

{\bf Introduction.} Gauge theories originated as an attempt by H. Weyl to use scale invariance to unify gravity and electromagnetism \cite{Weyl-1918}. But, their phenomenal impact did not commence until their migration from the domain of external space to that of internal degrees of freedom, like phase \cite{Weyl-1929}, but the name stuck. Invariance under gauge transformation, with non-Abelian generalizations \cite{Yang-Mills} has since explained electromagnetism, helped unify it with the weak nuclear force, and provided the framework for the strong nuclear force in quantum chromodynamics \cite{Aithchison-Hey}. Related concepts have also found applications in condensed matter systems \cite{Wen}.

Gauge structures were found to appear naturally in the adiabatic evolution of quantum systems, as definitively shown by Berry \cite{berry}, although anticipated in an earlier work on effective nuclear Hamiltonians in the Born-Oppenheimer approximation \cite{Mead-Truhlar}. This discovery led to multiple studies, both theoretical and experimental, that identified such structures in diverse phenomena such as nuclear magnetic resonance \cite{NMR_berry}, nuclear quadrupole resonance \cite{Tycko-NQR,Zee-PRA}, effective nuclear Hamiltonians in diatoms \cite{Moody-Wilczek-PRL-non-abelian} and in the context of molecular Kramers degeneracy \cite{Mead_Kramers-degeneracy} and in atomic collisions \cite{ZYGELMAN1987476}. In recent years, the same principle has been applied extensively to generate numerous phenomena based on synthetic gauge structures in neutral ultracold atoms \cite{Spielman-magnetic, Spielman-electric,RMP-Dalibard,Review-Goldman}.

\begin{figure}\includegraphics[width=\columnwidth]{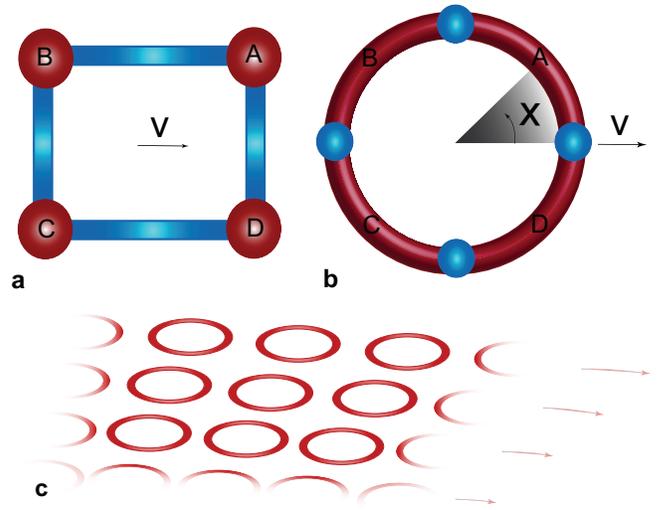}
\caption{{\bf Schematics of physical models.} {\bf a.} A literal rendering of the discrete model in real space with four lattice sites, labeled $A,B,C,D$ connected in a closed loop. {\bf b.} The topologically equivalent ring model used in our simulation with four wells separated by barriers; here $v$ is translational velocity and $x$ measures distance along the ring. {\bf c.} A lattice of such rings on a turntable can have a radial variation of velocities, to create a discrete gauge field in real space. }\label{schematic}
\end{figure}

Natural or synthetic, gauge structures remain typically associated with some internal degrees of freedom, in the latter case, the electronic and nuclear states of atoms, even when the parameters inducing the adiabatic changes could be external or even classical in nature. Notable exceptions include the impacts of Berry phase in  electronic orbital dynamics \cite{Niu-RMP-Berry-Phase}, however even there the carriers do not have independent existence beyond the crystalline lattice.  Furthermore, the dynamics involved in creating them has been less of interest than the resulting structures, partly due to the size and time scales involved. This is particularly so in the current context of ultracold atoms of interest in this paper. Here, we step beyond these constraints by demonstrating how gauge structures can be created in real physical space, where both the adiabatic parameters and most of the relevant degrees of freedom reside in external co-ordinate space. Specifically, the medium can be independent entities like individual atoms, or macroscopic quantum states like Bose-Einstein condensates (BEC) with gauge dynamics that can be directly imaged. Furthermore, with macroscopic quantum states, the time scales can be orders of magnitude larger than in electronic or nuclear processes, allowing direct access to the dynamics and even to freeze the evolution at specific instants. Since gravity is associated with general co-ordinate transformations \cite{Zee-gravity-book} but the other fundamental forces are tied to gauge transformations, the construction of gauge structures in macroscopic co-ordinate space may help to diminish that persistent gulf, with experiments that can probe both kinds of transformations in a shared space.

{\bf Gauge Structures in Adiabatic Evolution.} Synthetic gauge fields utilizes the identity of the mathematical structure that defines gauge freedom with that associated with geometric phase \cite{berry}. A quantum system, when evolved adiabatically to follow a specific eigenstate $|n(t)\rangle$ with energy $E(t)$ of the instantaneous Hamiltonian, acquires, apart from the dynamical phase $-\int_0^t dt E(t)/\hbar$, a geometric phase $ \vec{A}=i\langle n|\nabla|n\rangle$ as well, which depends only on the path, not the rate, of evolution. For an open path, the arbitrariness of reference gives the phase an inherent freedom which is removed for a closed path by the requirement of single-valuedness of quantum states. Specifically, under multiplication by a phase factor $|n\rangle \rightarrow e^{-i\chi}|n\rangle$, that freedom is reflected in the resulting transformation $ \vec{A}\rightarrow \vec{A}+\nabla \chi$, whereas integral over a closed path $\oint \vec{A}$ is fixed, and therefore by Stoke's theorem, so is $F=\nabla \times \vec{A}$. This is precisely the mathematical relation between the vector potential and the magnetic field, under an Abelian $U(1)$ gauge transformation. For two degenerate states, the adiabatic evolution restricted to their subspace leads to a $2\times 2$ matrix gauge potential, $ \vec{A}_{ ij}=i \langle \Phi_i|\nabla|\Phi_j\rangle$, that transform as components of a generally non-Abelian $U(2)=U(1)\times SU(2)$ gauge potential \cite{Wilczek-Zee-PRL-non-abelian}, with components of the field given by $F_{\mu\nu}=\partial_\mu A_\nu-\partial_\nu A_\mu- i[ A_\mu, A_\nu]$, where the indices label general parameters, and state indices are left out. It should be noted that the characteristics of gauge potential and gauge field identified here are also known as the Berry connection and Berry curvature \cite{Niu-RMP-Berry-Phase}.

\begin{figure}\includegraphics[width=\columnwidth]{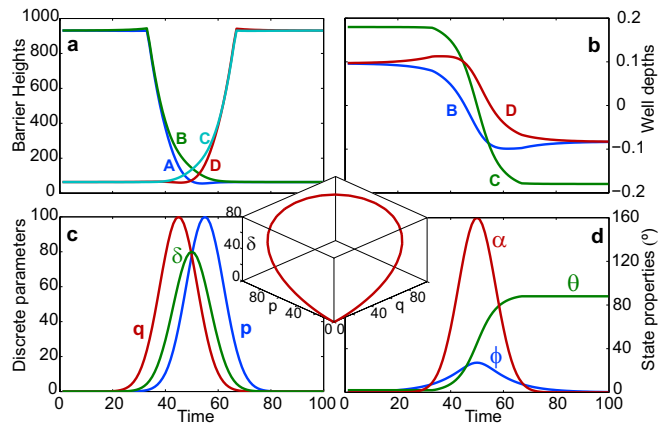}
\caption{{\bf Time evolution of the control parameters and properties.} Taking the depth of well A as reference, {\bf a.} the heights of the four barriers and {\bf b.} the depths of the remaining three wells are tracked as they are evolved in time. {\bf c.} The temporal path is set by choosing mutually offset Gaussian profile evolution for the underlying parameters $p,q$ and $\delta$ of the discrete Hamiltonian.  The inset shows their parametric locus. {\bf d.} The evolution of the eigenstates are characterized by the angles $\theta$ and $\phi$, obtained by re-parameterizing  $p,q$ and $\delta$ from the discrete Hamiltonian, along with the phase $\alpha$ the discrete counterpart of which evolves synchronously with $\delta$.}\label{parameters}
\end{figure}

In seeking implementations with cold atoms, eigenstates with zero eigenvalues were favored, as they eliminate unwanted dynamical phases. This established a link to the physics of STIRAP (Stimulated Raman Adiabatic Passage) \cite{STIRAP-RMP}, which uses such `dark' states to transfer population between atomic levels via an intermediate one without ever populating it. In separate developments, with advances in coherent manipulation of matter in nanostructures \cite{Ferry-Goodnick} and ultracold atoms \cite{Pethick-Smith}, it was noted that the same principle could be applied more dramatically to transport material particles across an intervening space without ever occupying it significantly \cite{Greentree,Eckert}.  In this paper, we bring these disparate ideas together for the purpose of constructing gauge structures in real physical space, involving translational degrees of freedom of the quantum states of massive particles.

{\bf Discrete Ring Model.} We start with a prototype discrete Hamiltonian which features (i) dark states, (ii) both Abelian and non-Abelian structures, (iii) complex coupling, (iv) one dimensional dynamics and (v) closed topology, which together will facilitate implementation and scalability in the continuum counterpart.  Models currently considered lack subsets of these features \cite{Fleischhauer-PRL-non-abelian-gauge,Spielman-ring,Review-Goldman,RMP-Dalibard}, typically requiring distinct configurations for different group structures and featuring open tree topology. Here we use an alternate four-level closed loop Hamiltonian:
\bn\label{Hamiltonian} H=
\left(
\begin{array}{cccc}
 0 & e^{i \alpha} p & 0 &  p \\
 e^{-i \alpha} p & \delta &  q & 0 \\
 0 &  q & 0 & e^{-i \alpha} q \\
 p & 0 & e^{i \alpha} q & \mp \delta
\end{array}
\right).\en
In the context of internal states of atoms, $p,q$ are transition amplitudes between the levels with $\alpha$ the associated phase, and $\delta$ the detuning of the laser coupling \cite{Das-wilson}. When $H_{44}=-\delta$, there are two degenerate dark states suitable for $U(2)$ implementation,
\bn\label{dark-states} \Phi_1&=&(e^{i\alpha} \cos\theta,0,-\sin\theta,0),\\
\Phi_2&=&(e^{i \alpha} \sin\phi \sin\theta,{\textstyle -\frac{1}{\sqrt{2}}}\cos\phi, \sin\phi \cos\theta,e^{i\alpha}{\textstyle \frac{1}{\sqrt{2}}}\cos\phi),\n\en
but when  $H_{44}=+\delta$, the degeneracy is lifted and the sole dark state $\Phi_1$ can be used for $U(1)$ implementation, while $\Phi_2$ corresponds to eigenvalue $\delta$. The coupling strengths $p,q$ and the detuning $\delta$ were reparameterized as
\bn\label{parametrization}
\Omega=\sqrt{\delta^2+2(p^2+q^2)},\h{2mm}\sin(\phi)=\frac{ \delta}{\Omega},\h{2mm}
\tan(\theta)=\frac{p}{q}.\en

If the system evolves slowly on the scale of the gap separating the dark state(s) from energetically adjacent states, the description can be confined to the subspace of dark states, and the state vector represented by them, $\Psi_i(t)=W_{ij}(t) \Phi_j(t)$, its index signifying the initial state $\Psi_i(0) = \Phi_i(0)$. Insertion in the Schr\"odinger equation yields coupled equations for the amplitudes,
\bn \label{evolution-equations} \dot{W}_{ij} =i \vec{A}_{ik} W_{kj}\cdot\dot{\vec{\mu}}, \h{5mm} \vec{A}_{ik}=i\langle \Phi_i|\vec{\nabla}|\Phi_k\rangle.\en
where $\vec{\mu}$ represents system parameters. The components of the gauge potential and the corresponding field are shown in Methods. Integration results in a path-ordered (${\cal P}$) integral for the evolution matrix, and when evaluated for a closed path in parameter space, $W_{\circ}={\cal P} e^{i\oint d\vec{\mu} \cdot \vec{A}}$, its trace ${\cal W}= {\rm tr}[W_{\circ}]$ is the gauge invariant generalization of the geometric phase factor, called the Wilson loop \cite{Wilson-PRD}. Its components can be made to correspond to the appropriate time-evolved quantum state \cite{Das-wilson}. Thus, the dependence on the gauge potential $\vec{A}$ shows that the synthetic gauge structures can be made manifest by time evolving the state in the space of parameters.

\begin{figure*}\includegraphics[width=\textwidth]{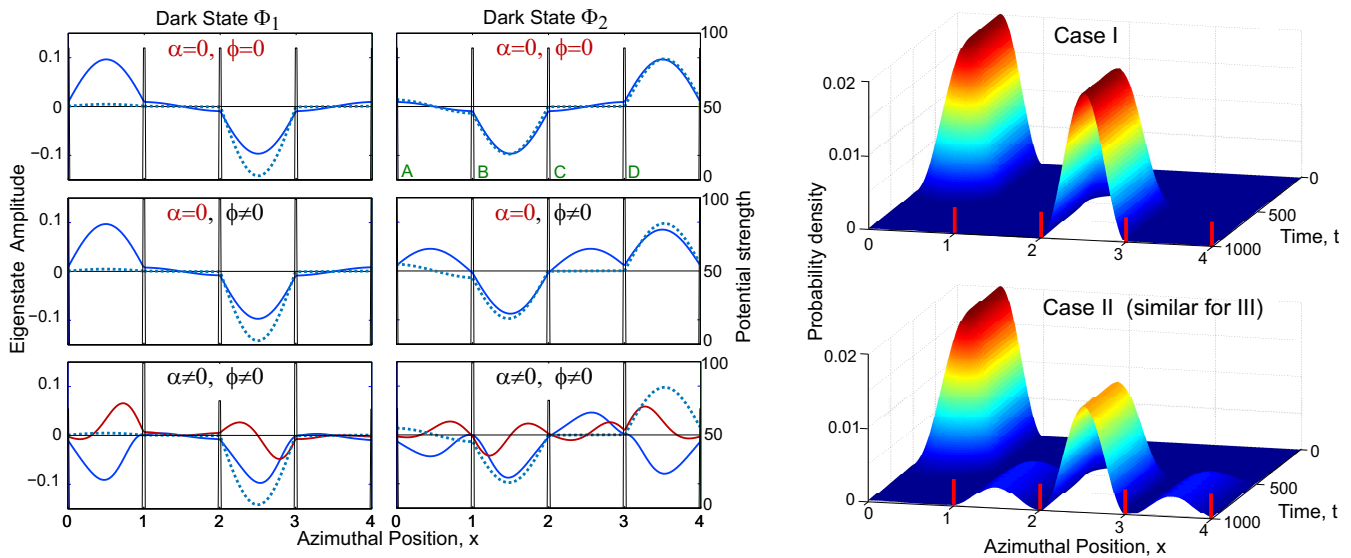}
\caption{{\bf The degenerate eigenstates and their time evolution.} (\emph{Line Plots}) The real (blue) and imaginary (red) components of the dark states, $\Phi_1$ on the left and $\Phi_2$ on the right panels, are shown for three cases that differ in regards to which of the parameters $\phi,$ and $\alpha$ are allowed to vary in time. The solid lines represent the states at the halfway point of the evolution, and the dashed lines at completion. The potential barriers are measured along the right axis, and share the label ($A,B,C,D$) of each right adjacent well. The last case, $\phi\neq 0,\alpha\neq 0$ is the only one with non-vanishing imaginary component which still vanishes at completion. (\emph{Surface Plots}) Time evolutions of the population in the wells for case (I) Abelian  (upper) and case (II) quasi-Abelian (lower) which is similar to that of case (III) non-Abelian. Red bars mark the barrier positions.}\label{eigenstates}
\end{figure*}

{\bf Continuum Ring Model.} The Hamiltonian can be mapped to a system of four coupled potential wells, arranged in a closed loop, identified as $A,B,C,D$ in Fig.~\ref{schematic}(a). Each well corresponds to a bare atomic level. Two elements of the discrete Hamiltonian have ready equivalents: The onsite energies provide the detunings, $\pm \delta$, and the strengths of the barriers separating the wells measure the inverse of the coupling strengths, $p,q$. However, the couplings are real-valued, therefore a complex Hamiltonian such as in Eq.~(1) does not naturally arise in the continuum counterpart. We overcome that by allowing the system to translate, with velocity $v$, as shown in Fig.~\ref{schematic}(a). This assures that equal and opposite phases are acquired by the wavefuction in evolving through couplings $A{\small \leftrightarrow} B$ and $C{\small \leftrightarrow} D$ that parallel the direction of translation, while couplings $B{\small \leftrightarrow} C$ and $D{\small \leftrightarrow}A$ orthogonal to the translation remain unaffected, just as required by the structure of the Hamiltonian (\ref{Hamiltonian}).

For our simulations, we tweak this literal implementation, into a topologically equivalent, but smooth, circular configuration, shown in Fig.~\ref{schematic}(b), that avoids sharp corners but contains all the essential physics.  The state is confined to the azimuth of a toroidal trapping potential \cite{Phillips_Campbell_hysteresis}, with sufficiently strong confinement along its minor axis to populate only the lowest transverse eigenstate, so the dynamics is effectively one-dimensional. The four wells are created by uniformly spaced barrier potentials, which we assume to be rectangular, the precise shape not critical as long as the specific well structure is created and adequate number of parameters are available \cite{GOD-1}.

We now address the challenges of how to transcribe the discrete model into the continuous ring system, while keeping the gauge structures intact.  While the discrete Hamiltonian has three effective parameters $\theta,\phi$ and $\alpha$, the continuum model offers the complications of a much greater set, hence we seek a manageable subset that is yet sufficient for our purpose.  The inter-well couplings are affected by shape and size of the barriers, but by keeping the shape and width fixed, we can characterize them each by a single parameter, the barrier height.  Likewise, by keeping the barriers uniformly spaced, each well can be defined by its depth. Thus, our simulations can be restricted to eight Hamiltonian parameters $\{{\cal H}_i|i\in 1,\cdots,8\}$ that include the velocity , $v$ of the ring, the four barrier heights and three of the well depths (the fourth serves as energy reference).

In the discrete model, the analytical structure of the eigenvalues and eigenstates is preserved as long as the Hamiltonian retains its form, so the time evolution is simply a matter of varying the three parameters along a desired path. It is impractical to do the same in the continuum model, because the interdependency of a greater number of available parameters makes it non-trivial to predict how the eigenvalues and eigenstates will change as the Hamiltonian parameters are varied. Therefore, the key to successful implementation is to focus instead on the eigenstates themselves. We choose the path of evolution to maintain the relevant features of the dark states and ensure the wavefunction remains projected as desired onto their subspace at all times, and then, adjust the Hamiltonian accordingly. Towards that, we need to identify eight appropriate state parameters $\{{\cal S}_i|i\in 1,\cdots,8\}$ to map to the Hamiltonian parameters.

{\bf External Physical Parameters.}  While the discrete case has only four eigenstates, the ring system has infinite, but adiabatic evolution permits the system to be described by the lowest band of four eigenstates, which we label $\Phi_i, {i\in 0,1,2,3}$, and the corresponding eigenvalues $E_i$, so the counterparts of the discrete dark states conveniently remain $\Phi_1$ and $\Phi_2$. For each well, $\sigma=A,B,C,D$, the well centers are designated with a subscript $\sigma_0$, and the populations of the states are $n_{i(\sigma)}=\int dx |\Phi_i(x)|^2$ integrated between centers of adjacent barriers.

We use the structure of the eigenstates in Eq.~(\ref{dark-states}) to define parameters that are close equivalents of the discrete parameters as follows $\mu\equiv\{\theta,\phi,\alpha\}$,
\bn \theta\equiv{\cal S}_1&=&\sin^{-1}\left(\sqrt{n_{1(C)}/\{n_{1(A)}+n_{1(C)}\}}\right)\n\\  \phi\equiv{\cal S}_2&=&\tan^{-1}\left({\rm Re}\{\Phi_2(C_0)\}/{\rm Re}\{-\sqrt{2}\cos({\cal S}_1)\Phi_2(B_0)\}\right)\n\\
\alpha\equiv{\cal S}_3&=&{\rm arg}\{\Phi_1(A_0)\}\en
The definition of $\theta$ follows from the expression for the state $\Phi_1$ since all the population for that state is confined to wells $A$ and $C$. Likewise $\phi$ is defined by the expression for the state $\Phi_2$ specifically the amplitudes in wells $B$ and $C$, the amplitudes used rather than the population to capture the sign of the $\sin(\phi)$.  The global phase for the two dark states are set by choosing negative values for the amplitudes in well B and well C for states $\Phi_1$ and $\Phi_2$ respectively. The phase $\alpha$ is derived from the first component of $\Phi_1$ in Eq.~(\ref{dark-states}). It is directly tied to the velocity component along the azimuth $v\sin(2\pi x/L)$, x measuring distance along the ring circumference up to $L=4$.
Since the vector sum of the velocity components around the ring vanishes, this is consistent with the discrete model, where the dark state structure requires the net phase of the coupling terms add up to multiples of $2\pi$. Five additional properties are picked that utilizes the five additional degrees of freedom available in the Hamiltonian (see Methods) to optimize the evolution.

{\bf Time evolution.} For the purpose of evolving the state to manifest the gauge structures, as mentioned in Eq.~(\ref{evolution-equations}), we choose a path based on varying the parameters in the underlying discrete model with Gaussian profiles \cite{Bergmann-PRA-STIRAP-geometric},
\bn h=h_0e^{-(t-\tau_h)^2/\sigma_h^2}\h{1cm} h\in\{p,q,\delta,\alpha\}.\en
All pulse widths are set to be $\sigma_h=10$, relative delays to satisfy $\tau_p-\tau_\delta=\tau_\delta-\tau_q=\tau$, $\tau_\alpha=\tau_\delta$ and amplitudes $p_0=q_0=100$, $\delta_0=80$ and $\alpha_0=8\pi/9$ rad $=160^\circ$. The profiles for $p,q,\delta$ are plotted in Fig.~\ref{parameters}(c), $\alpha$ matches the evolution of $\delta$ but with a different scale. The inset shows the curve they trace out in $3D$.  The corresponding variations for the parameters $\theta,\phi,\alpha$ are shown in Fig.~\ref{parameters}(d).  The resulting evolution of the Hamiltonian parameters is shown in Fig.~\ref{parameters}(a,b). The effectiveness of the mapping ${\cal H} \rightarrow {\cal S}$ in maintaining the functional form of the dark states is confirmed in Fig.~\ref{eigenstates}, where their snapshots are plotted at the middle and at the end of the evolution, for the three separate cases described below. Specifically, $\Phi_1$ maintains minimal population in wells B and D, while $\Phi_2$ has equal population in them.

Our results can be best understood in the context of the eigenstates in Eq.~(\ref{dark-states}). We start with all the population in well A.  Since the evolution is such that $\phi=0$, and $\theta \sim 0$, the initial state coincides with the dark state $\Phi_1$.  At termination, $\phi=0$ once again but $\theta \sim \pi/2$, so well A is unoccupied, and any population in well C corresponds to $\Phi_1$ while any population in wells $B$ and $D$ correspond to $\Phi_2$. We now show that three distinct classes of gauge structures can be implemented in the four well ring.

\begin{figure}\includegraphics[width=\columnwidth]{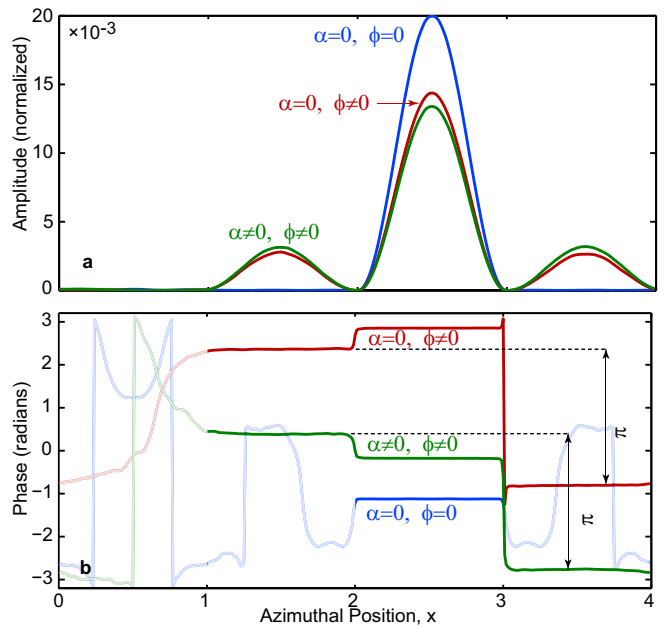}
\caption{{\bf The final state at completion of time evolution.} {\bf a.}  Magnitude and {\bf b.} phase of the final states for the three cases considered: (I) Abelian $\phi=\alpha=0$ (blue), (II) quasi-Abelian $\phi\neq 0, \alpha=0$ (red) (III) non-Abelian $\phi\neq 0, \alpha\neq 0$ (green). For (II) and (III) the magnitude vanishes in well $A$ while for (I), in wells $A,B$ and $D$, hence in {\bf b} the phase is masked in those wells as irrelevant. The phase difference of $\pi$ between wells B and C for cases (II) and (III) highlights agreement with the analytical structure of the dark states. }\label{final-state}
\end{figure}

{\bf Signatures of Gauge Structures.} \\
(I) \emph{Abelian} ($\phi=\alpha=0$): In this case there is no coupling between the two dark states, as is evident from Eq.~(\ref{dark-states}), and the system effectively follows $\Phi_1$.  Confined to a subspace of one state, this corresponds to an Abelian gauge structure. The time evolution in Fig.~\ref{eigenstates} confirms that the population always remains in state $\Phi_1$. Specifically, at termination, the entire population has transferred from well A to well C, as seen clearly in Fig.~\ref{final-state}.  As such, the only relevant phase is in well C, where it has a constant value, and since the dynamical phase is maintained at zero, this arises form geometric phase which measures the $U(1)$ gauge rotation.

(II) \emph{Quasi-Abelian} ($\phi\neq 0, \alpha=0$): Now, the two states are coupled, so even though the system starts off in state $\Phi_1$, Fig.~\ref{eigenstates} shows that $\Phi_2$ acquires population during evolution, and in contrast to case (I), at termination the population is distributed among wells B,C, and D, indicating a superposition of the two dark states. This mixing of states is a measure of the $U(2)$ gauge rotation. Figure~\ref{final-state} shows that there is a phase difference of $\pi$ between wells B and D as predicted by the sign difference in Eq.~(\ref{dark-states}) between the respective components. However, this case is still effectively Abelian, despite the degeneracy, because there is only a single field component which commutes for any pair of points in the parameter space, and furthermore the gauge invariant Wilson Loop computed for multiple closed paths with a common starting point is independent of the order of the loops (see Methods).

(III) \emph{Non-Abelian} ($\phi\neq 0, \alpha\neq 0$): As in case (II) there is coupling between the two states, and the population evolves similarly and is finally distributed among wells B,C and D. However, as Fig.~\ref{final-state} shows, the presence of non-vanishing $\alpha$ due to translation clearly impacts both the magnitude and the phase.  While there is a visible difference with case (II) in magnitude, it is the difference in the phase between the two cases which is striking. This is significant, for it makes the corresponding $U(2)$ rotation here truly non-Abelian, as supported by the non-commutation of the field components and that the Wilson loop computed over multiple closed paths sharing a starting point can be shown to depend on their order (see Methods) \cite{Das-wilson}. Nevertheless, since $\alpha$ vanishes at termination, the phase difference of $\pi$ between wells B and D at the end is found to be maintained as in case (II).

{\bf Discussion and Conclusions.}
Our considerations can be generalized to create gauge fields in real space. Each ring can be viewed as mimicking an atom. Thus, a multiplicity of them distributed in space, with each acquiring a different gauge structure based on its position, would be the analog of a discrete gauge field. One way to implement this is shown schematically in Fig.~\ref{schematic}(c), where an array of such rings are arranged on a rotating turntable. The translational velocity will then vary with radius, and therefore, so will the non-Abelian gauge transformation as in case (III) above. Towards this, a two-dimensional lattice could in principle be adapted, using additional potentials to create a super-lattice structure such that individual plaquettes of four sites could mimic closed rings. Alternately, a vertical stack of ring shaped lattices could be creating by intersecting multiple sheets of light with counter-propagating Laguerre-Gaussian beams carrying orbital angular momentum \cite{Vaity_LG-lattice}. If a collective coherent state like a Bose-Einstein condensate \cite{Pethick-Smith} is used, the size and time scales can be macroscopic, and the dynamics of the quantum state evolution, analogous to that of internal atomic states that create synthetic gauge fields, can be then directly observed, mapped and even frozen during evolution.

\noindent{\bf Methods}

\noindent{\bf Gauge Structure.} The gauge structures associated with the discrete Hamiltonian in Eq.~(\ref{Hamiltonian}) are briefly summarized here. For a single dark state, the gauge potential and the gauge field both have only one component
\bn A_{\alpha}&=&i \langle \Phi_1|\partial_{\alpha}|\Phi_1\rangle=-\cos^2\theta\n\\
F_{\theta\alpha}&=&\partial_\theta A_\alpha-\partial_\alpha A_\theta= \sin(2\theta)\en
For two dark states, the components of the potential are $2\times2$ matrices, represented here in terms of the four generators of the $U(2)$ gauge group, $I_2$ the identity and $\sigma_{i=x,y,z}$ the Pauli spin matrices, along with the projection operators $\sigma_{\uparrow(\downarrow)}=\frac{1}{2}(I_2\pm \sigma_z)$. The non-zero components are
\bn  A_\theta&=&-\sin\phi\ \sigma_y, \h{5mm} \\
  A_\alpha&=&-{\textstyle \frac{1}{2}}\sin\phi \sin(2\theta) \sigma_x\n\\&&-\cos^2(\theta)\sigma_\uparrow
-{\textstyle \frac{1}{2}}[1- \sin^2(\phi)\cos(2\theta)]\sigma_\downarrow \n
\en
Likewise, the gauge field $F_{\mu\nu}=\partial_\mu A_\nu-\partial_\nu A_\mu- i[ A_\mu, A_\nu]$ has the following non vanishing contributions from the curl and the commutator in the chosen basis
\bn \label{curls}\partial_\theta A_\phi-\partial_\phi A_\theta\h{-1mm}&=&\h{-1mm}-\cos\phi\ \sigma_y\n\\
\partial_\phi A_\alpha-\partial_\alpha A_\phi\h{-1mm}&=&\h{-1mm}-{\textstyle\frac{1}{2}}\cos\phi \sin(2\theta)\sigma_x+{\textstyle\frac{1}{2}}\sin(2\phi)\cos(2\theta) \sigma_\downarrow\n\\
 \partial_\theta A_\alpha-\partial_\alpha A_\theta\h{-1mm}&=&\h{-1mm} \sin(2\theta) [\sigma_\uparrow\h{-1mm} -\sin^2\h{-1mm} \phi\ \sigma_\downarrow]-\sin\phi\cos(2\theta)\sigma_x,\n\\
-i[ A_\theta, A_\alpha]
&=&{\textstyle \frac{1}{2}}\sin(\phi)\cos(2\theta)[1+\sin^2(\phi)]\sigma_x\n\\&&  -\sin^2\phi \sin(2\theta)\sigma_z.\en
Notably, when $\alpha=0$, all commutators vanish, and both the gauge potential and field have only one non-vanishing component, both proportional to a single generator, $A_\theta\propto\sigma_y$ and $F_{\theta\phi}\propto\sigma_y$, assuring that they would commute for any pair of points in the parameter space. The covariance of the field $F\rightarrow UFU^\dagger$ under gauge transformation $U$ guarantees that the commutator of the field would vanish in every gauge, and the gauge structure is therefore effectively Abelian despite the degeneracy. This is further confirmed by the Wilson loop matrix which can be analytically computed in this case, $W_\circ(\Lambda)=I_2\cos(\Lambda)+i\sigma_y\sin(\Lambda)$ where $\Lambda=-\oint d\theta\sin(\phi)$ and it is clear that for any two closed paths, $A,B$, $[W_\circ(\Lambda_A),W_\circ(\Lambda_B)]=0$, hence also the commutator of their traces, the Wilson loops.

\noindent{\bf Optimization Parameters.} Having defined the primary properties corresponding to the discrete model, we now define five additional properties that utilizes the five additional degrees of freedom available in the Hamiltonian. They are chosen to ensure the desired time evolution, and all of them are defined so that their optimal values are identically zero at all times. With the ground state energy $E_0$ chosen as the energy reference, two of the parameters are the value of a dark state energy ${\cal S}_4=E_1$  and their energy separation.
${\cal S}_5=(E_2-E_1)/E_1$, minimizing which would maintain the required  constancy and degeneracy, the former of course only up to variation in $E_0$ itself. The key features of the eigenstates are maintained with the following parameters,
\bn &&{\cal S}_6=n_{2(B)}-n_{2(D)}\n\\
&&{\cal S}_7={\rm Re}\{\Phi_1(D_0)/{\rm sgn}\{\Phi_1(A_0)\}\},\n\\
&&
 {\cal S}_8={\rm Re}\{\Phi_1(B_0)/{\rm sgn}\{\Phi_1(C_0)\}\}.\en
The parameter ${\cal S}_6=0$ ensures that the population of the dark state $\Phi_2$ in wells B and D are identical. The last two parameters ensure that the dark state $\Phi_1$ has a node at the centers of wells $B$ and $D$ to approximate the zero population in those wells.  Furthermore, their phases are matched to those at the centers of  wells A and C in order to implement the condition that there is vanishing phase shift between wells A and D and between wells B and C respectively as evident from the Hamiltonian in Eq.~(\ref{Hamiltonian}).

There are two points worth noting. First, the constraint  $\theta\in[2^\circ,88^\circ]$ is imposed, relevant towards the start and end of the evolution when $p\rightarrow 0$ and  $q\rightarrow 0$  respectively. This is necessary because if $\theta \rightarrow 0^\circ$ or $\theta \rightarrow 90^\circ$, the degeneracy of the eigenvalues of the dark states becomes almost exact and the Jacobian matrix becomes singular. Secondly, although the path closes in the $p,q,\delta$ space, the parameter $\theta$ clearly does not return to the original value, so the Hamiltonian evolution based on the angular parameters, does not return to the original configuration.  However, the path can be closed by a rapid projection of the Hamiltonian to the original configuration, and we found the essential features remain qualitatively intact.

\noindent{\bf Path Optimization.}
For a desired evolution of the eigenstates, the path can be chosen in the space ${\cal S}$ which, for the purpose of numerical simulation, is mapped by a discrete set of points that can be associated with time steps $\{t_n\}$ for some chosen rate of traversal. At each point, a trial vector ${\cal H}^{(1)}$ is picked based upon values at the previous point, which would yield a non-optimal state vector ${\cal S}^{(1)}$ that deviates by $\delta {\cal S}^{(1)}$ from the point on the path ${\cal S}(t_n)$. Solution of the system of linear equations $\delta {\cal S}_j^{(1)}= \sum_i J_{ji} \delta {\cal H}_i^{(1)}$ where $J_{ji}=\frac{\partial {\cal S}_j}{\partial {\cal H}_i}$ is the Jacobian matrix, then yields the necessary adjustments $\delta {\cal H}^{(1)}$ for an improved vector ${\cal H}^{(2)}={\cal H}^{(1)}+\delta {\cal H}^{(1)}$ with reduced deviations $\delta{\cal S}^{(2)}$. This is iterated until until a set ${\cal H}(t_n)\equiv{\cal H}^{(k)}$ is found to yield the vector ${\cal S}(t_n)$ within set tolerances.

%

\vspace{5mm}
{\bf Acknowledgments.}
\begin{acknowledgments} K.K.D. acknowledges the support of NSF under Grants No. PHY-1313871,
No. PHY-1707878, and No. PHY11-25915, and the KITP at
UCSB where this work began under a Kavli Scholarship. \end{acknowledgments}

\vspace{5mm}
{\bf Author contributions.}\\
KKD generated the essential ideas in the paper, wrote it and did all the analytical calculations.  MG wrote most of computer codes used for simulations.

\vspace{1cm}
{\bf Competing interests.}\\
The authors declare that they have no competing interests.
\vfill

\vspace{1cm}
{\bf Corresponding Author.}\\
Correspondence to das@kutztown.edu.
\vfill
\end{document}